\documentclass[12pt]{article}
\usepackage{amsmath}
\usepackage{epsfig}
 \usepackage{graphics}
\usepackage{latexsym}
\usepackage [ansinew]{inputenc}
\usepackage {hyperref}
\begin{document}

\vskip 2cm
\title{  Logarithmic correction to scaling for multi-spin strings in the $AdS_5$
black hole  background }
\author{\\					  
A.L. Larsen${}^{*}$
}
\maketitle
\noindent
{\em
Department of Physics and Chemistry, University of Southern Denmark,
Campusvej 55, 5230 Odense M,
Denmark}
\begin{abstract}
\baselineskip=1.5em
\hspace*{-6mm}
We find new explicit solutions describing closed strings spinning with equal angular momentum 
in two independent planes in the $AdS_5$ black hole spacetime. These are $2n$ folded strings 
in the radial direction and also winding $m$ times around an angular direction. We
especially consider these solutions in the long string and high temperature limit, where it
is shown that there is a logarithmic correction to the scaling between energy and spin.
This is similar to the one-spin case. The strings are spinning, or actually orbiting around 
the black hole of the $AdS_5$ black hole spacetime, similarly to solutions previously found in black hole 
spacetimes. \end{abstract}	  \vskip 1cm\noindent
 \vskip 1cm\noindent
$^{*}$Electronic address: all@ifk.sdu.dk\\

 \section{Introduction}
 There has been a lot of progress concerning the conjectured duality \cite{maldacena,gubser,witten}
 between super string theory in $AdS_5\times S^5$ and ${\cal {N}}=4$ SU(N) super Yang-Mills 
 theory in Minkowski space, beyond the super gravity limit. One step forward was obtained 
 by Berenstein, Maldacena and Nastase \cite{bmn}, where one considers a particular limit of $AdS_5\times S^5$, 
 where it reduces to a plane wave, in which string theory is exactly solvable.   
 Another step forward was obtained by Gubser, Klebanov and Polyakov \cite{klebanov}, where it was observed that the first
  subleading term in the scaling relation between energy and spin for a rigidly rotating 
  one-spin string in $AdS_5$ is logarithmic	  (actually, spinning strings \cite{inigo}, 
  as well as pulsating strings \cite{de vega}, in $AdS$ space were originally studied more
  than 10 years ago) 
  \begin{equation}
 E-S\sim \ln(S)
\label{eq1}
\end{equation} 
 in	the long string limit, 
 similarly to results previously found for certain 
  Yang-Mills operators \cite{gross,georgi,floratos, korchemsky, dolan}	  at the gauge theory side.

It was soon realized that most progress came from considering multi-spin strings 
in $AdS_5\times S^5$. However, most papers on multi-spin strings deal with strings where
most of the spins are in the $S^5$ part of the background (in general there are two angular momenta in the $AdS_5$ 
part and three in the $S^5$ part), since the computation of energies at the gauge 
theory side then reduces to a study of certain spin chains \cite{minahan}. Moreover, some
 non-renormalization results have been obtained
for large angular momenta in $S^5$; see for instance \cite{hofman,tseytlin3}
(and references given therein). 	 Many papers take therefore only the time coordinate, 
and the corresponding conserved energy, from $AdS_5$ and up to three angular momenta in 
$S^5$, while others take only a single spin in $AdS_5$. 

There have only been few papers 
with explicit examples of multi-spin strings in the $AdS_5$ part of the background 
\cite{tseytlin3,tseytlin8,alex}. In \cite{tseytlin3} (footnote number 9) it was claimed that there cannot be any 
two-spin string
solutions with $\sigma$ depending  $r$ coordinate. However, they only considered
$2$-folded strings. These strings gave rise to a powerlaw relationship between energy and 
spin, thus no logarithmic relationship!  In
 \cite{alex}, a generalization of the string leading to equation (\ref{eq1}) was obtained. 
 Namely, a two-spin string with two equal angular momenta in two independent planes 
of $AdS_5$, leading to the result 
\begin{equation}
 E-2S\sim \ln(S)
\label{eq2}
\end{equation}	   in	the long string limit.	 These strings are $2n$-folded with $n>2$. 

Also backgrounds related to $AdS_5$, in one way or the other, has been considered. 
Of interest to this letter, especially the $AdS_5$ black hole background, which is supposed
to describe $5$ dimensional $AdS$ spacetime at finite temperature \cite{witten1}. A one-spin string 
orbiting around the black hole was constructed in \cite{armoni}, leading in the long string and high 
temperature limit to a result similar to equation (\ref{eq1}). A two-spin string was constructed in 
\cite{mosaffa}, but it was not extended in the radial direction and it did not lead to a relation
similar to equation (\ref{eq2}); instead a powerlaw relationship was again obtained.  

The purpose of the present letter is to construct a two-spin string solution in the $AdS_5$
black hole background, which is extended in both the radial and an angular direction,
and which leads to a scaling relation of the form of equation (\ref{eq2}). 
	The claim of reference \cite{tseytlin3}, that there are no two-spin string	 solutions with $\sigma$ 
	depending $r$ coordinate, holds true in this case also, but we shall again show that 
	by allowing many foldings, such solutions are indeed possible.   These strings will be 
	orbiting around the black hole, rather than spinning around their 
	 center of mass, similarly to solutions previously found in black 
	 hole spacetimes \cite{inigo,armoni,kar,hendy}.
  	  \section{The two-spin string solution}
We have the
$AdS_5$ black hole
line element \cite{witten1}
\begin{eqnarray}
ds^2&=&-(
1+H^2r^2-M/r^2)dt^2+\frac{dr^2}{1+H^2r^2-M/r^2}  \nonumber   \\
&+&r^2(d\beta^2+\sin^2\beta
d\phi^2+\cos^2\beta d\tilde{\phi}^2 )
\label{eq3}
\end{eqnarray}	  where $M$ is proportional to the mass, 
which for large mass is proportional to $T^4$, where $T$ is the temperature \cite{witten1}. 
The horizon is located at
\begin{eqnarray}
r_h^2=\frac{-1+\sqrt{1+4MH^2}}{2H^2}
\label{eq4}
\end{eqnarray} We are interested in strings which are extended in the $r$ and $\beta$ 
directions, and which have two identical angular momenta in the $\phi$ and 
$\tilde{\phi}$ directions.  
 For our ansatz
\begin{equation}
{X}^\mu=(c_0\tau,r(\sigma),\beta(\sigma),\omega \tau ,\omega \tau)
\label{eq5}
\end{equation}
where
$(c_0, \omega) $
  are constants, the tangent vectors are
\begin{equation}
\dot{X}^\mu=( c_0,0,0,\omega,\omega),\ \ \ X'^\mu=(0,r',\beta',0,0)
\label{eq6}
\end{equation}	    where dot and prime denote derivative with respect to $\tau$ and 
$\sigma$, respectively. 
The 5 equations of motion  (we use the orthonormal gauge) 
\begin{equation}
\ddot{X}^\mu-X''^\mu+\Gamma^\mu_{\rho\sigma}(\dot{X}^\rho\dot{X}^\sigma-
X'^\rho
X'^\sigma)=0
\label{eq7}
\end{equation}
lead to
\begin{eqnarray}
 &r''-\frac{H^2r+M/r^3}{1+H^2r^2-M/r^2}r'^2-(1+H^2r^2-M/r^2)(H^2r  +M/r^3) c_0^2 &\nonumber\\
	 &-(1+H^2r^2-M/r^2)r \beta'^2+(1+H^2r^2-M/r^2)r\omega^2=0 &
 	 \label{eq8}
	 \end{eqnarray}
 	 \begin{eqnarray}
\beta'=
 \frac{k}{r^2}&
 \label{eq9}
 \end{eqnarray}
  where $k$ is an integration  constant.
  The induced metric on the world-sheet is
\begin{eqnarray}
g_{\tau\sigma}&=&G_{\mu\nu}\dot{X}^\mu X'^\nu=0 \label{eq10}\\
g_{\tau\tau}&=&G_{\mu\nu}\dot{X}^\mu\dot{X}^\nu= -(1+H^2r^2-M/r^2)c_0^2+
r^2\omega^2 \label{eq11}\\
   g_{\sigma\sigma}&=&G_{\mu\nu}{X'}^\mu
X'^\nu=  (1+H^2r^2-M/r^2)^{-1}r'^2+  r^2  {\beta'}^2 \label{eq12}
  \end{eqnarray}
Then the orthonormal gauge  constraints $g_{\tau \sigma}=0$ and  $g_{\tau \tau}+g_{\sigma\sigma
}=0$, lead to
\begin{eqnarray}
 r'^2&=&  (
 1+H^2r^2-M/r^2
 )\left( (1+H^2r^2-M/r^2)c_0^2   -r^2\omega^2- \frac{k^2}{r^2}\right)\nonumber
\\& =&
 \frac{ 1+H^2r^2-M/r^2}{r^2}(\omega^2-H^2c_0^2) (r_+^2-r^2)(r^2-r_-^2)
 \label{eq13}
 \end{eqnarray}
where

\begin{eqnarray}
r^2_\pm=\frac{1}{2
(\omega^2-H^2c_0^2)
}\left( c_0^2\pm\sqrt{c_0^4+4 (H^2c_0^2-\omega^2)(Mc_0^2+k^2)}\right)
\label{eq14} 
\end{eqnarray}  Notice that (\ref{eq13}) is the integral of (\ref{eq8}), so the complete dynamics is 
	given by (\ref{eq9}) and (\ref{eq13}).  
  We are interested in solutions which are oscillating 
  (in terms of the spatial world-sheet coordinate $\sigma$)  between positive
$r_-$ and $r_+$,
  thus we demand
\begin{eqnarray}
\omega^2>H^2c_0^2
\label{eq15}  
\end{eqnarray}
   and
  \begin{eqnarray}
c_0^4+4 (H^2c_0^2-\omega^2)(Mc_0^2+k^2)  >0
\label{eq16}   
\end{eqnarray}	  which are equivalent to \begin{eqnarray}
0<	\frac{\omega^2}{ H^2c_0^2}-1< \frac{
c_0^ 2}{  4H^2(Mc_0^2+k^2)  } 
\label{eq17}	     
\end{eqnarray}
By Taylor expansion of (\ref{eq14}), we find that 
   \begin{eqnarray}
 r_-^2\geq M+k^2/c_0^2
 \end{eqnarray}
 Comparing with (\ref{eq4}), we conclude that $r_->r_h$, i.e. 
 the string is always orbiting outside the horizon.  
 
Imposing the boundary conditions $r(0) =r_+$ and $\beta(0) =0$, we
have the solutions  
  \begin{eqnarray}
  \sigma=-\frac{1}{\sqrt{\omega^2-H^2c_0^2}}\int_{r_+}^r
\frac{xdx}{\sqrt{ (1+H^2x^2-M/x^2)(r_+^2-x^2)(x^2-r_-^2)}}
\label{eq18}
\end{eqnarray}
  \begin{eqnarray}
 \beta(\sigma) =k\int_0^{ \sigma}\frac{dx}{r^2(x)}
 \label{eq19} 
 \end{eqnarray}
  These solutions hold for non-constant $r(\sigma)$. 
  The solution for constant $r$ is given in \cite{mosaffa}. In this letter  we consider strings which are $2n$ folded in the $r$ direction and
winding $m$ times around the $\beta $ direction, 
thus the string is closed and consists of $2n$ segments.
 We  then have the periodicity conditions $r(\sigma+2\pi/n) =r(\sigma)$ and
 $ \beta(\sigma+2\pi) =2\pi m+  \beta(\sigma)$. The first condition becomes

 \begin{eqnarray}
  \pi=
   \frac{n}{\sqrt{\omega^2-H^2c_0^2}}\int_{r_-}^{r_+}
\frac{xdx}{\sqrt{ (1+H^2x^2-M/x^2)(r_+^2-x^2)(x^2-r_-^2)}}
\label{eq20}
\end{eqnarray}
  while the second leads to

  \begin{eqnarray}
   2\pi m
 = \frac{2kn}{\sqrt{\omega^2-H^2c_0^2}}\int_{r_-}^{r_+}
\frac{dx}{x\sqrt{ (1+H^2x^2-M/x^2)(r_+^2-x^2)(x^2-r_-^2)}}
\label{eq21}
  \end{eqnarray}
  For fixed $M$ and $H$, we have 3 free parameters namely $c_0$, $\omega$
and $k$,
  but 2 of them are determined by the two previous equations.
  Then we have also the 2 integers $n$ and $m$. Notice that $n$ and $m$ may be constrained 
  since $k$, $c_0$ and $\omega$ must be real. For instance, for $M=0$ we find that $n/m>2$ \cite{alex}. 

 Notice that when the bounds in equations (\ref{eq15}) and (\ref{eq16}) are saturated, the corresponding strings 
 are infinitely long respectively pointlike in the $r$-direction. In the latter case, 
 the periodicity condition (\ref{eq20}) for $r$ obviously drops out. For these pointlike 
 strings (in the $r$-direction), we refer to \cite{mosaffa}. 

The conserved energy is given by
 \begin{eqnarray}
 E=\frac{c_0}{2\pi\alpha'}
 \int_0^{2\pi}
 (1+H^2r^2(\sigma )-M/r^2(\sigma))d\sigma
 \label{eq22}
 \end{eqnarray}
Similarly, the conserved spins are
\begin{eqnarray}
S_1=\frac{\omega}{2\pi\alpha'}
\int_0^{2\pi}r^2(\sigma )\cos^2\beta (\sigma
)d\sigma, \ \ \ \
  S_2=\frac{\omega}{2\pi\alpha'}\int_0^{2\pi}r^2(\sigma )\sin^2\beta
(\sigma )d\sigma
\label{eq23}
  \end{eqnarray}
 In our case, $S_1=S_2\equiv S=(S_1+S_2)/2$.  Then  we get

\begin{eqnarray}
   E=\frac{nc_0}{\pi\alpha'\sqrt{\omega^2-H^2c_0^2}}
   \int_{r_-}^{r_+}
  xdx\sqrt{ \frac{1+H^2x^2-M/x^2}{(r_+^2-x^2)(x^2-r_-^2)}}
  \label{eq24}
  \end{eqnarray}

\begin{eqnarray}
S=\frac{n\omega}{2\pi\alpha'\sqrt{\omega^2-H^2c_0^2}}
   \int_{r_-}^{r_+} \frac{x^3dx}{\sqrt{
(1+H^2x^2-M/x^2)(r_+^2-x^2)(x^2-r_-^2)}}
\label{eq25}
\end{eqnarray}  \section{Analysis of the solution}
Now introduce
\begin{eqnarray}
r_0 ^2=\frac{1+\sqrt{1+4MH^2}}{2H^2}
\label{eq26}
\end{eqnarray} 
 Then equations (\ref{eq20}), (\ref{eq21}), (\ref{eq24}) and (\ref{eq25}) lead to

 \begin{eqnarray}
  \pi=
   \frac{n}{ H\sqrt{\omega^2-H^2c_0^2}}\int_{r_-}^{r_+}
\frac{x^2dx}{\sqrt{ (x^2+r_0^2)( x^2-r_h^2)(r_+^2-x^2)(x^2-r_-^2)}}
\label{eq27}
\end{eqnarray}

  \begin{eqnarray}
   2\pi m
  = \frac{2kn}{H\sqrt{\omega^2-H^2c_0^2}}\int_{r_-}^{r_+}
\frac{dx}{\sqrt{ (x^2+r_0^2)(x^2-r_h^2)(r_+^2-x^2)(x^2-r_-^2)}}
\label{eq28}
\end{eqnarray}
\begin{eqnarray}
   E=\frac{nHc_0}{\pi\alpha'\sqrt{\omega^2-H^2c_0^2}}
   \int_{r_-}^{r_+}
  dx\sqrt{ \frac{(x^2+r_0^2)(x^2-r_h^2)}{(r_+^2-x^2)(x^2-r_-^2)}}
  \label{eq29}
  \end{eqnarray}
 \begin{eqnarray}
S=\frac{n\omega}{2\pi H\alpha'\sqrt{\omega^2-H^2c_0^2}}
   \int_{r_-}^{r_+} \frac{x^4dx}{\sqrt{
(x^2+r_0^2)(x^2-r_h^2)(r_+^2-x^2)(x^2-r_-^2)}}
\label{eq30}
\end{eqnarray}
In the special  case $M=0$,
   there is no black hole and we have pure $AdS_5$.  Then the integrals (\ref{eq27})-(\ref{eq30}) reduce to 
   the results obtained in \cite{alex}, as they should. 
Now introduce   the dimensionless constants
 \begin{eqnarray}
y_0=H^2r_0^2,\ \ \ \
y_h=H^2r_h^2,\ \ \ \
y_-=H^2r_-^2,\ \ \ \
y_+=H^2r_+^2   
\label{eq31}
\end{eqnarray}
as well as the new dimensionless integration variable
\begin{eqnarray}
 y=H^2x^2
 \label{eq32}
 \end{eqnarray}
Performing changes of integration variable, similarly to the Appendix of \cite{armoni}, the 
integrals 
(\ref{eq27})-(\ref{eq30}) lead to 
  \begin{eqnarray}
\pi = \frac{n}{2 \sqrt{ y_+ - y_- }
 \sqrt{ \omega^2 - H^2 c^2_0 }}
\int ^1 _0 \frac{ \sqrt{
t + \frac{ y_- }{y_+ - y_- }} dt}{ \sqrt{ \left( t +
\frac{ y_- +y_0 }{y_+ - y_- } \right) \left( t +
\frac{y_- - y_h }{y_+ - y_-} \right) \left( 1 - t
\right) t }}
\label{eq33}
\end{eqnarray}
	\begin{eqnarray}
2 \pi m = \frac{ H^2kn }{ \left( y_+ - y_- \right) ^{\frac{3}{2}}
\sqrt{ \omega ^2 - H^2 c^2_0 }}
\int
^1 _0 \frac{ dt}{ \sqrt{ \left( t + \frac{ y_- +y_0 }{y_+
- y_-} \right) \left( t +
\frac{y_- -  y_h }{y_+ - y_- } \right) \left( 1-t \right)  t \left(  t +
\frac{y_-}{y_+ - y_-} \right) }}
\label{eq34}
\end{eqnarray}
	  \begin{eqnarray}
E= \frac{nc_0\sqrt{ y_+ - y_- } 
 }{2 \pi  \alpha ' \sqrt{
\omega ^2 - H^2 c^2_0}}
\int ^1 _0
\sqrt{ \frac{ \left( t + \frac{y_- +y_0 }{y_+ - y_-}
\right)  \left( t + \frac{ y_- - y_h }{y_+ - y_- }
\right) }{ \left( t+ \frac{y_- }{y_+ -y_-} \right)
\left( 1- t
\right) t}} dt
\label{35}
\end{eqnarray}
\begin{eqnarray}
S =  \frac{n \omega\sqrt{ y_+ - y_- }
 }{4 \pi H^2 \alpha ' \sqrt{\omega^2 -H^2c^2_0}}
 \int ^1 _0
\frac{ \left( t+ \frac{y_- }{y_+ - y_-} \right) ^{
\frac{3}{2}} dt}{ \sqrt{ \left( t + \frac{y_- +y_0
}{y_+ -y_-} \right) \left(t + \frac{y_- - y_h }{y_+ -y_-
} \right) \left( 1-t \right) t }}
\label{eq36}
\end{eqnarray}	
 The integrals (\ref{eq33})-(\ref{eq36}) are of hyperelliptic type, and cannot be directly evaluated. 
 However, we are only interested in the long string limit. 
 In the long string limit we find from equation (\ref{eq14}) 
	 \begin{eqnarray}
\omega/Hc_0=1+2\eta, \ \ \  \ \eta<<1
\label{eq37}
\end{eqnarray}
Then 
	 \begin{eqnarray}
\sqrt{\omega^2-H^2c_0^2}\approx 2Hc_0\sqrt{\eta}
\label{eq38}
\end{eqnarray}

	 \begin{eqnarray}
y_+\approx 1/4\eta
\label{eq39}
\end{eqnarray}
	 \begin{eqnarray}
y_-\approx H^2(M+k^2/c_0^2)
\label{eq40}
\end{eqnarray}
We furthermore take the high temperature limit. 
Long strings and high temperature means that $y_+>>y_->>1$. 
Then also $  y_->>(y_h, \ y_0) $, and we can approximate  
	 \begin{eqnarray}
(t+\frac{y_-+y_0}{y_+-y_-})^{-1/2}  
\approx
\left(t+\frac{y_-}{y_+-y_- }\right)^{-1/2}
\left(1+\sum_{j=1}\frac{(2j-1)!!\left(  \frac{-y_0}{y_+-y_-}  \right)^j}
{ (2j)!!\left(t+\frac{y_-}{y_+-y_- }\right)^{j} }\right)   
\label{eq41}
\end{eqnarray}
\begin{eqnarray}
(t+\frac{y_--y_h}{y_+-y_-})^{-1/2}   \approx \left(t+\frac{y_-}{y_+-y_- }\right)^{-1/2}\left(1+\sum_{j=1}\frac{ (2j-1)!!\left( \frac{y_h}{y_+-y_-}  \right)^j}
 { (2j)!!\left(t+\frac{y_-}{y_+-y_- }\right)^{j} }\right)   
 \label{eq42}
 \end{eqnarray}  and similarly for the powers $+1/2$ appearing in (\ref{35}).
This can be inserted into (\ref{eq33})-(\ref{eq36}), and then the integrals reduce to elliptic integrals, which can be
evaluated to all orders in our approximation. 
Approximating also the elliptic integrals \cite{abr}, we get  

			 \begin{eqnarray}
  \pi = \frac{n}{2H c_0} \left( \ln \left(\frac{1}{4\eta H^2 \left(M  + k^2/c^2_0\right)}\right)
  +{\cal O}(\eta,\ 1/MH^2)\right)
  \label{eq43}
  \end{eqnarray}
				 \begin{eqnarray}
2 \pi m &=& \frac{ H  kn}{c_0} \left( \frac{2}{ H^2\left( M+k^2/c^2_0  \right) } 
	  +{\cal O}(\eta,\ 1/M^2H^4)\right)
	  \label{eq44}
   \end{eqnarray}
	\begin{eqnarray}
E &=& \frac{n }{4 \pi \alpha ^{\prime} H \eta} \left(
1+\eta\ln \left(\frac{1}{ 4 \eta H^2 \left(
M+ k^2/c^2_0\right)} \right)
	  +{\cal O}(\eta)\right)
	  \label{eq45}
	  \end{eqnarray}	
	  \begin{eqnarray}
S &=& \frac{n }{8 \pi \alpha ^{\prime} H^2 \eta} \left(
1-\eta\ln \left(\frac{1}{ 4 \eta H^2 \left(
M+ k^2/c^2_0\right)} \right)
	  +{\cal O}(\eta)\right)
	  \label{eq46}
	  \end{eqnarray}	 where ${\cal O }(\eta,\ 1/MH^2)$ means ${\cal O}(\eta )$ 
or ${\cal O}( 1/MH^2)$.  From (\ref{eq43}) and (\ref{eq44}), we get $Hc_0\approx\omega$ and $H^2k$
 				 \begin{eqnarray}
Hc_0\approx\omega\approx \frac{n}{2 \pi } \ln \left( \frac{1}{ 4 \eta
\left( \frac{ n^2}{2 \pi ^2 m^2} \pm \frac{n}{ \pi m} \sqrt{ \frac{
n^2}{4 \pi ^2 m^2}- H^2 M} \right) } \right)
\label{eq47}
\end{eqnarray}
\begin{eqnarray}
H^2k \approx\frac{n}{2 \pi }\left( \frac{n }{ 2 \pi m} \pm \sqrt{ \frac{n^2}{4 \pi ^2 m^2}- H^2 M} \right)
 \ln \left( \frac{1}{ 4 \eta
\left( \frac{ n^2}{2 \pi ^2 m^2} \pm \frac{n}{ \pi m} \sqrt{ \frac{n^2}{4 \pi ^2 m^2}- H^2 M} \right) } \right) 
\label{eq48}
\end{eqnarray}	  Thus we get the condition that $n/m$ should be relatively large 
\begin{eqnarray}	 n/m\geq 2\pi H\sqrt{M} \label{eq49} \end{eqnarray}
Let us stress again that there was a claim in reference \cite{tseytlin3}  (footnote number 9), 
although in the $M=0$ case, that there are no two-spin string solutions with $\sigma$ depending $r$ 
coordinate, and thereby no solutions leading to a logarithmic dependence between energy and 
spin in the long string limit. Here we have shown that by allowing a large number of 
foldings, given by (\ref{eq49}), such solutions are indeed possible. It should be stressed that
the bound (\ref{eq49}) holds only in the long string and high temperature limit. For instance, in the opposite limit 
where the string is not extended in the $r$ direction, and where the periodicity condition 
(\ref{eq20}) drops out, we have instead the bound \cite{mosaffa}

 \begin{eqnarray}
m\leq \frac{c_0}{4\sqrt{M}}
\end{eqnarray}
(in our notation), where $c_0$ is a free parameter for fixed $M$.   From (\ref{eq45}) and (\ref{eq46}) we get   
\begin{eqnarray}
{E}/{H}- 2S \approx \frac{n }{2 \pi \alpha ^{\prime}
H^2}\ln\left( \frac{1}{4 \eta H^2 \left( M +
k^2/ c^2_0\right)}  \right)
\label{eq50}
\end{eqnarray}
	 i.e.	 \begin{eqnarray}
E/H-2S \approx\frac{n}{2\pi\alpha'H^2} \ln\left(\frac{S\alpha'}{M+k^2/c^2_0}\right)
\label{eq51}
\end{eqnarray}
  which is equivalent to \begin{eqnarray}
{E}/{H}-2S \approx \frac{n}{2\pi H^2 \alpha ^{\prime}
} \ln \left( \frac{ H^2  \alpha
^{\prime}S}{\frac{n^2}{2 \pi ^2 m^2} \pm \frac{n}{\pi m}
\sqrt{ \frac{n^2}{4 \pi ^2 m^2} - H^2M }}\right)
\label{eq52}
\end{eqnarray}
which can also be expressed in terms of the t'Hooft coupling 
$\lambda=(H^2\alpha')^{-2}$.  

Notice that we get two solutions, as in the case where
the string is not extended 
in the radial direction \cite{mosaffa}. In our case we get  from equations 
(\ref{eq39}) and (\ref{eq40})
		 \begin{eqnarray}
r_+^2=\frac{1}{4H^2\eta	 } \label{eq53} \end{eqnarray}
whereas 
	  \begin{eqnarray}
r_-^2=\frac{1}{H^2}\left( \frac{ n^2}{2 \pi ^2 m^2} \pm \frac{n}{ \pi m}
\sqrt{ \frac{n^2}{4 \pi ^2 m^2}- H^2 M} \right)
\label{eq54}
\end{eqnarray}
\section{Conclusion}  	 In conclusion, we have shown that it is possible to obtain a logarithmic correction to the 
	 scaling between energy and spin for a two-spin string orbiting around the black hole in 
	 the $AdS_5$ black hole background, contrary to a claim in the literature 
	 \cite{tseytlin3} (footnote number 9).  
	
  In order to obtain this result, we must take the long
  string and the high temperature limit, similarly to the one-spin case. The two-spin string 
  consists of $2n$ segments and is winding $m$ times around an angular direction.
Furthermore, we must allow a large number of foldings so that $n/m$ is of the order of 
$H\sqrt{M}$. The result of getting a logarithmic correction to the scaling between energy and 
spin, gives hope to finding the dual operator at the gauge theory side along the lines of
\cite{klebanov}. This is however out of the scope of this letter.  

  It is easy to generalize these solutions to the $AdS_5$ black hole times $S^5$, by introducing up
to three angular momenta (R-charges) in $S^5$.

\vskip 24pt
{\bf Acknowledgements}:\\
I would like to thank A. Albertsen for secretarial help.

\end{document}